\documentstyle[12pt,epsf]{article}

\newfont{\Bbb}{msbm10 scaled 1200}     
\newcommand{\mathbb}[1]{\mbox{\Bbb #1}}

\newcounter{apps}

\newcounter{prs}[section]

\newcounter{cors}

\newcounter{figs}

\newcounter{th}

\newcommand{\be}{\begin{equation}}
\newcommand{\ee}{\end{equation}}
\newcommand{\bea}{\begin{eqnarray*}}
\newcommand{\eea}{\end{eqnarray*}}
\newcommand{\beaa}{\begin{eqnarray}}
\newcommand{\eeaa}{\end{eqnarray}}

\newcommand{\ba}{\begin{array}}
\newcommand{\ea}{\end{array}}

\newcommand{\lb}{\label}
\newcommand{\g}{\gamma}

\newcommand{\ra}{\rightarrow}

\newcommand{\td}{\tilde}

\newcommand{\bt}{\beta}
\newcommand{\p}{\partial}

\newcommand{\Ld}{\Lambda}
\newcommand{\vp}{\varphi}

\newcommand{\HH}{{\mathcal{H}}}
\newcommand{\Hd}{{\mathcal{H}}^*}

\begin{document}

\begin{flushright}
ITEP-TH-08/04\\
\end{flushright}

\medskip

\begin{center}
\bigskip
{\large\bf Non RG logarithms via RG equations}

\bigskip
\bigskip

Dmitry Malyshev%
\footnote{
dmalyshe@princeton.edu}

\bigskip

\bigskip

\bigskip
{February 2004}

\end{center}

\bigskip
\bigskip

\begin{abstract}
We compute complete leading logarithms in $\Phi^4$ theory
with the help of Connes and Kreimer RG equations.
These equations are defined in the Lie algebra dual to
the Hopf algebra of graphs.
The results are compared with calculations in parquet
approximation.
An interpretation of the new RG equations is
discussed.
\end{abstract}


\section{Introduction}

One of the most subtle problems in QFT is the divergence of
Feynmen integrals.
This problem was solved by Bogolubov and Parasuk \cite{Bog}.
The process of consistent subtraction of divergences is called
Bogolubov's R-operation.
Recently Connes and Kreimer discovered that the R-operation has
the structure of Hopf algebra of graphs \cite{K1,CK1,CK2}.

The existence of the Hopf algebra has many interesting theoretical
consequences such as connection to deformation quantization \cite{Ionescu:2003nm}
and noncommutative geometry \cite{Connes:1998qv}.
From the practical point of view the Hopf algebra sheds light on
combinatorics of R-operation and simplifies the
multiloop calculations \cite{Kreimer:1999wh,Broadhurst:1999ys}.
The Hopf algebra is also tightly connected to the renormalization
group in QFT \cite{CK2,GMS}.
In the present paper we utilize  this connection for the
calculation of the leading logarithms in $\Phi^4$ theory.
We will show that the Connes and Kreimer RG equations
are more restrictive and
enable one to find the leading logarithms in an arbitrary point
(at least in some theories).
The ordinary methods give the asymptotics only in
symmetric points with respect to external momenta because
in arbitrary points there are logarithms of ratios of
external momenta invisible for RG equations.

The main object of consideration is the linear space of graphs
$\HH$.
This space is infinite dimensional with basis labeled by graphs.
Dual space $\Hd$ is the space of linear functions on $\HH$.
Any QFT may be interpreted as a vector $F\in\Hd$.
If we know the couplings, then the function $F$ maps the graphs
to the corresponding Feynman integrals.

RG equations describe the change of couplings with the change of
the scale.
The set of couplings defines a vector
$F\in\Hd$.
The change of $F$ is described by RG equations in the space $\Hd$.
These equations have the form of linear
differential equations.
The transition from the ordinary RG equations
to the RG equations in the space $\Hd$
seems to be trivial but it is a crucial step,
because the last equations show up to be more informative.
Indeed, a linear equation on a vector is a system of equations on
the coordinates.
In the case of graphs it means that there is an RG equation for
any Feynman integral.
What could be an interpretation of the new RG
equations?
The problem is that different Feynman diagrams with the same
number of vertices have approximately the same magnitude
and it seems that they are indistinguishable in experiments.
But this is not the case when we have several particles with the
same interaction but with different charges.
In this case, Feynman integrals may give different contributions in
different observables, and we can distinguish the integrals by
comparing the observables.
At the end we will consider an example of a theory with two
charges.

The paper is organized as follows.
In the next section we will state the generalized RG equations and
discuss their properties.
In section 3 we calculate the leading logarithms using the RG
equations.
In section 4 we compare the results with the calculations in
parquet approximation.
After that we study the RG equations in the model with two
charges.

\section{Generalized RG equations}

In this section, we describe the RG equations in the linear space
of graphs.
The Hopf algebra of graphs appeared as a mathematical
structure underlying the R-operation \cite{CK1}.
The Hopf algebra is dual to a Lie algebra of graphs.
The RG equations are generated by some special element of the Lie
algebra.
This element is called the beta-function \cite{CK2}.
In the present paper we will focus on the application of RG
equations to the case of leading logarithms
in massless $\Phi^4$ theory in $d=4$.

Throughout the paper, we will consider only
1 particle irreducible diagrams
with four external legs.
Let $\g_n$ be an $n$-loop diagram.
We define $F(\g_n)$ to be the leading logarithm contribution to
the diagram.

The ordinary RG equation (for the leading logarithms) is
\be
\frac{\p}{\p\log\Ld^2}F^{(n)}=\bt(g)\frac{\p}{\p g}F^{(n-1)},
\ee
where $\bt(g)$ is the one-loop beta-function and
$F^{(n)}$ is the $n$ loop contribution to the four point
function
\be
F^{(n)}=\sum_{\g_n}\frac{g^{n+1}}{S_{\g_n}}F({\g_n}),
\ee
where $S_{\g_n}$ is the symmetry factor for the graph $\g_n$.

The RG equation in the space of linear functions of graphs
is an equation on the vector $F\in\Hd$.
The one-loop RG equation for the coordinates of this vector is
\be\lb{rgmain}
\frac{\p}{\p\log\Ld^2}F(\g_n)=\sum_{\g_{n-1}=\g_n/\g_1}F(\g_{n-1}),
\ee
as before $F(\g_n)$ is the leading logarithm contribution for the
graph $\g_n$.
The symbol $\g_n/\g_1$ denotes the graph obtained by the contraction
of subgraph $\g_1\subset\g_n$ into a point.
The sum is over $(n-1)$-loop graphs obtained by the contraction of
one-loop subgraphs in $\g_n$.
Note that in this equation we have no numerical factors,
such as symmetry factors $S_{\g_n}$.
For the explanation of 'miracle' cancellation of these factors see
\cite{ll4}.

Equation (\ref{rgmain}) has two remarkable properties: it
preserves the orientation of the graphs and it is linear in $F$.

Let us define the orientation in $\Phi^4$ theory for the graphs
contributing to the leading logarithms.
All such graphs are 2 particle reducible \cite{dyatlov},
i.e. in any graph there are 2 lines
such that the cutting of these lines makes the graph disjoint.
Let $q$ denote the sum of the momenta flowing through the cut
lines.
Then $q^2$ is one of the three Mandelstam variables
$s$, $t$ or $u$.
Consequently there are three possible orientations of the graphs
which we will denote by the same letters $s$, $t$ and $u$.
The conservation of the orientation means that the graphs on
both sides of equation (\ref{rgmain}) have the same orientation.

The linearity of equation (\ref{rgmain}) enables one to write it
in the form of geodesic equation.
We define the operator
\be
\bt^{-1}\g_n=\sum_{\g_{n-1}=\g_n/\g_1}\g_{n-1},
\ee
here $\g_n$, $\g_{n-1}$ denote the basis vectors in $\Hd$
corresponding to graphs $\g_n$, $\g_{n-1}$.
Recall that the operator $\bt$ acts by insertion of one-loop
subgraphs.
The operator $\bt^{-1}$ acts in the opposite way by
contraction of one-loop subgraphs, though strictly speaking it is
not the inverse to $\bt$.
The non trivial components of $\bt^{-1}$ are
\be
(\bt^{-1})_{\g_n}^{\g_{n-1}}
=\left[
\ba{l}
1, \;\;if\;\; \g_{n-1}=\g_{n}/\g_{1};\\
0, \;\;if\;\; \g_{n-1}\neq\g_{n}/\g_{1}.
\ea
\right.
\ee
Equation (\ref{rgmain}) takes the form
\be
\frac{\p}{\p\tau}F-F\circ\bt^{-1}=0,
\ee
where $\tau=\log\Ld^2$.
In the components, this equation reads
(we assume the summation over $\g_{n-1}$)
\be
\frac{\p}{\p\tau}F_{\g_n}-(\bt^{-1})_{\g_n}^{\g_{n-1}}F_{\g_{n-1}}=0,
\ee
here $F_{\g_n}:=F(\g_n)$.
We see that in the space $\Hd$ the operator $\bt^{-1}$ is a
connection associated with the differentiation over $\tau=\log\Ld^2$.

\section{Leading logarithms}

In this section, we find the leading logarithms in
$\Phi^4$ theory.
The RG equations exist in any theory but the possibility to find
the non RG logarithms is specific for the $\Phi^4$ theory, because in
the derivation we will use some specific properties of graphs.
In this paper, we will not consider the Sudakov double logarithms
and infrared divergences.

The derivation is divided into three steps.
First, we consider the graphs which depend only on one external
momentum, then the graphs depending on two momenta, and at the end
we consider the general case.

\begin{figure}[h]
\begin{center}
\leavevmode
\epsfxsize 250pt
\epsffile{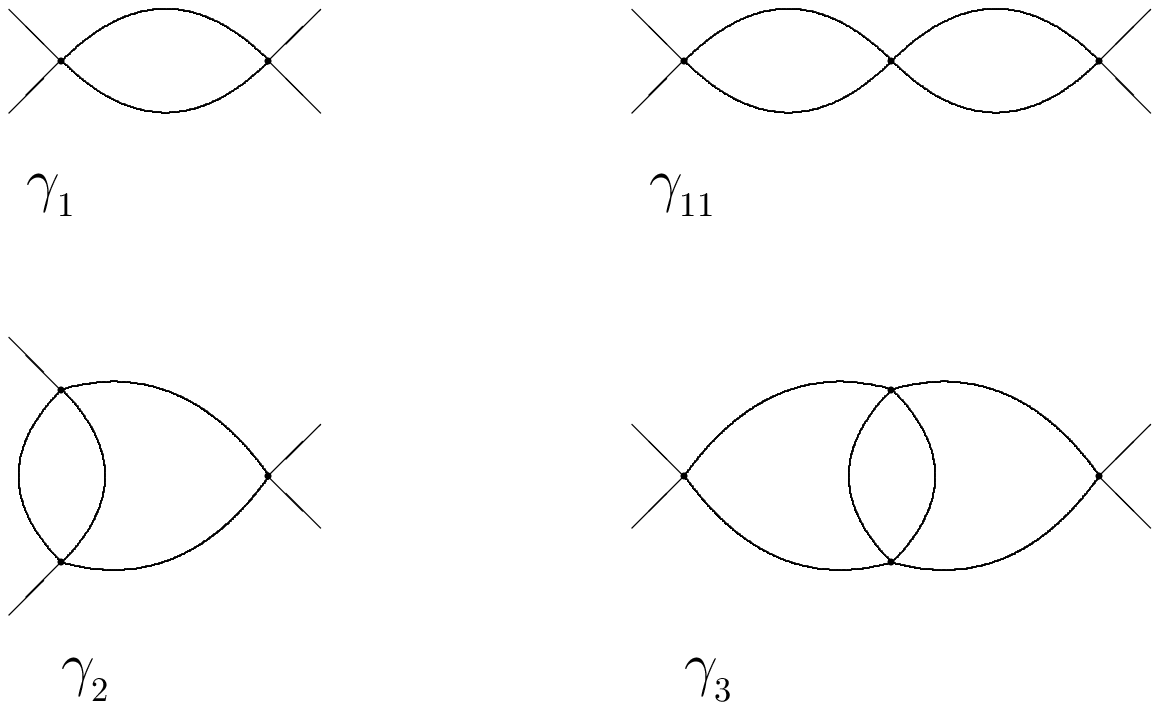}
\caption{\bf Notations for graphs}
\end{center}
\end{figure}

1. Let a graph $\g_n$ depend only on one external momentum $q$.
Then the only dimensionless combination is $q^2/\Ld^2$, and the
leading logarithm has the form
\be
F(\g_n)=c(\g_n)\left(\log\frac{\Ld^2}{q^2}\right)^n.
\ee
We find the leading logarithms for such graphs by induction.
The result for the one-loop graph is
\be
F(\g_1)=\log\frac{\Ld^2}{q^2}.
\ee
Now we use equation (\ref{rgmain}) in order to express
the coefficient $c(\g_n)$ in terms of the
coefficients $c(\g_{n-1})$ for some $(n-1)$-loop graphs.
The answer is as follows
\be\lb{1mom}
c(\g_n)=\frac{1}{n}\sum_{\g_{n-1}=\g_n/\g_1}c(\g_{n-1}).
\ee
If we consider a symmetric point in external momenta,
then this formula is valid for any graph $\g_n$ \cite{ll4}.
For an arbitrary point, we also have to find
the coefficients in front of the non RG logarithms.
As an example of application of formula (\ref{1mom}), we find the
coefficients for the graphs $\g_{11}$ and $\g_3$.
In the graph $\g_{11}$ we can contract the left and the right
one-loop subgraphs; each time the result of the subtraction is the
graph $\g_1$
\be
c(\g_{11})=\frac{1}{2}(c(\g_1)+c(\g_1))=1.
\ee
In the graph $\g_3$ we can contract only the one-loop subgraph in
the middle $\g_3/\g_1=\g_{11}$
\be
\lb{ll3}
c(\g_{3})=\frac{1}{3}c(\g_{11})=\frac{1}{3}.
\ee
The leading logarithms are
\beaa
F(\g_{11})&=&\left(\log\frac{\Ld^2}{q^2}\right)^2;\\
F(\g_3)&=&\frac{1}{3}\left(\log\frac{\Ld^2}{q^2}\right)^3.
\eeaa

2. The next step is to consider the graphs that depend on two external
momenta.
We may assume that $p>>q$.
(In the case $p<<q$ we can put $p=0$ and we get the previous case
of a graph depending on one external momentum.)

In the case of two external momenta
the leading logarithms have the form
\be
F(\g_n)=\sum_{k=0}^n c_k(\g_n)\left(\log\frac{\Ld^2}{q^2}\right)^{n-k}
\left(\log\frac{p^2}{q^2}\right)^k.
\ee
If we know the leading logarithms for $(n-1)$-loop graphs, then we
can use formula (\ref{rgmain}) and find the first $n$ coefficients
$c_0,\ldots,c_{n-1}$.
The problem is with the last term
\be
 c_n(\g_n)\left(\log\frac{p^2}{q^2}\right)^n,
\ee
which vanishes as we differentiate with respect to $\log\Ld^2$.

In order to find this last coefficient,
we note that the integral over $p$ is equal to some $(n+1)$-loop
diagram which depends only on $q$
\be
\int \frac{d^4p}{p^2(p+q)^2}F(\g_n;p,q)=F(\g_{n+1};q).
\ee
In this equation, we know the right hand side, since it depends
only on one external momentum.

The integral is effectively
\be
\int\frac{d^4p}{p^2(p+q)^2}=\int_0^{\log\Ld^2/q^2}d\log\frac{\Ld^2}{p^2}.
\ee
After the integration we get the relation
\be\lb{inte}
\sum_{k=0}^n\frac{c_k(\g_n)}{k+1}=c(\g_{n+1}).
\ee
As an example, we find the leading logarithms for the graph $\g_2$.
\be
F(\g_2)=c_0\left(\log\frac{\Ld^2}{q^2}\right)^2
+c_1\left(\log\frac{\Ld^2}{q^2}\right)\left(\log\frac{p^2}{q^2}\right)
+c_2\left(\log\frac{p^2}{q^2}\right)^2.
\ee
First we use equation (\ref{rgmain}) in order to find $c_0$ and
$c_1$
\be
\frac{\p}{\p\log\Ld^2}F(\g_2)=F(\g_1)=\log\frac{\Ld^2}{q^2},
\ee
consequently
\bea
c_0&=&\frac{1}{2};\\
c_1&=&0.
\eea
The integration over $p$ gives the graph $\g_3$ with
$c(\g_3)=1/3$ (see equation (\ref{ll3})).
Now we use (\ref{inte}) in order to find $c_2$
\be
\frac{c_2}{3}+\frac{c_1}{2}+\frac{c_0}{1}=c(\g_3).
\ee
The result is
\be
c_2=-\frac{1}{2}.
\ee
The leading logarithms are
\beaa
F(\g_2)
&=&\frac{1}{2}\left(\log\frac{\Ld^2}{q^2}\right)^2
-\frac{1}{2}\left(\log\frac{p^2}{q^2}\right)^2\\
\lb{2rg}
&=&\left(\log\frac{\Ld^2}{q^2}\right)\left(\log\frac{\Ld^2}{p^2}\right)
-\frac{1}{2}\left(\log\frac{\Ld^2}{p^2}\right)^2.
\eeaa

3. The calculation in the general case is similar to the case of
two external momenta.
But now we have $(n+1)(n+2)/2$ coefficients for an $n$-loop
diagram.
Among these coefficients, $n+1$ correspond to terms without $\log\Ld^2$
and are not fixed by RG equations.
Integration over one of the external momenta gives an $n+1$-loop
diagram with two external momenta.
Since we already know the result
for such diagrams, we can find the unknown $n+1$
coefficients.
The only subtle point here is to prove that the
expressions in front of the coefficients do not vanish,
i.e. that the analogue of equation (\ref{inte}) is not degenerate and we can
find all $n+1$ coefficients.

We see that the leading logarithms in $\Phi^4$ theory may be
calculated with the help of Connes and Kreimer RG equations.
The logic of this derivation may be reversed and
we can calculate the leading logarithms using parquet
approximation in order to check the new RG equations.

\section{Parquet approximation}

The computation of the leading logarithms in $\Phi^4$ theory was
performed by A.M.Polyakov in the appendix of the work \cite{polyakov}.
The diagrams contributing to the leading logarithms have
two-particle cross sections with three possible orientations \cite{dyatlov}.
Consequently, the vertex function has the following
representation \cite{polyakov}

\begin{figure}[h]
\begin{center}
\leavevmode
\epsfxsize 350pt
\epsffile{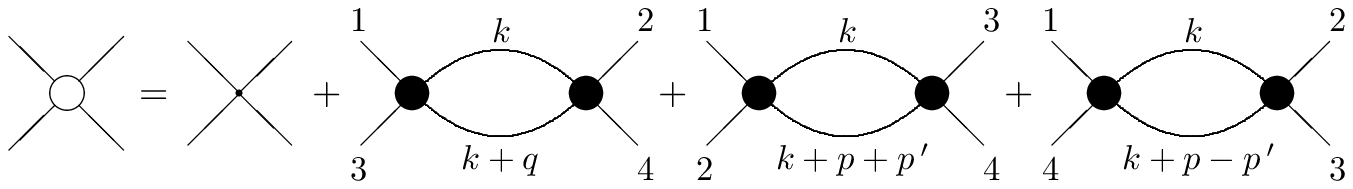}
\caption{\bf Four-point vertex function}
\end{center}
\end{figure}

The integration momentum $k$ is chosen to be
smaller than any other integration momentum in the shaded blocks.
The external momenta are defined as follows
\be
p_{1,3}=q/2\pm p\;,\;\;\;\;\;\;\;\;
p_{2,4}=-q/2\pm p\,',
\ee
so that $s=(p+p\,')^2$, $t=q^2$ and $u=(p-p\,')^2$.
We let $p\,'>>p>>q$ and denote
\be
\xi=\log\frac{\Ld^2}{{p\,'}^2},\;\;\;
\zeta=\log\frac{\Ld^2}{{p\,}^2},\;\;\;
\eta=\log\frac{\Ld^2}{q^2}.
\ee
This representation corresponds to the integral equation
\beaa\nonumber
F(\xi,\,\zeta,\,\eta)=-g
+3\int_0^\xi F^2(x,\,x,\,x)dx
+\int_\xi^\zeta F(x,\,x,\,x)F(\xi,\,x,\,x)dx\\
\lb{inteq}
+\int_\zeta^\eta F(\zeta,\,x,\,x)F(\xi,\,x,\,x)dx,
\eeaa
where the diagrams in $s$ and $u$ channels (the last two terms)
correspond to
$$
F_s(\xi)+F_u(\xi)=2\int_0^\xi F^2(x,\,x,\,x)dx
$$
and the diagram in $t$ channel (the second term) gives
\beaa
\nonumber
F_t(\xi,\,\zeta,\,\eta)=
\int_0^\xi F^2(x,\,x,\,x)dx
+\int_\xi^\zeta F(x,\,x,\,x)F(\xi,\,x,\,x)dx\\
\lb{tchan}
+\int_\zeta^\eta F(\zeta,\,x,\,x)F(\xi,\,x,\,x)dx.
\eeaa
The solution of equation (\ref{inteq}) gives the summation of leading
logarithms.
We can use this equation in order to find the leading logarithms
for individual Feynman integrals.
We expand the vertex function in diagrams and substitute the
expansion in the equation.
Then we can find the leading logarithms recursively, since an
$n$-loop diagram on the left hand side will be equal to some
integrals of diagrams with no more than $(n-1)$-loops.
For example, consider the diagram $\g_2$ in $t$ channel
that depends only on $q$ and $p$, i.e. on $\eta$ and $\zeta$.
Using equation (\ref{tchan}), we find that this diagram
equals
\be
F_{\g_2}(\zeta,\,\eta)
=\int_0^\zeta F_{\g_1}(x,\,x,\,x)F_{\g_0}dx
+\int_\zeta^\eta F_{\g_1}(\zeta,\,x,\,x)F_{\g_0}dx,
\ee
where $F_{\g_1}$ and $F_{\g_0}$ correspond to one-loop and
zero-loop diagrams, i.e. we insert the one-loop diagram in place of the
left shaded block and a vertex instead of the right shaded block.
The result is
\be
F_{\g_2}(\zeta,\,\eta)
=\int_0^\zeta xdx +\int_\zeta^\eta \zeta dx=\zeta\eta-\frac{1}{2}\zeta^2,
\ee
and we see that this result coincides with formula (\ref{2rg})
obtained from the RG equations.
Another example is the diagram $\g_3$ in $t$ channel.
This diagram has two possible decompositions: we can represent it
as the diagram $\g_2$ in place of the left block and the vertex
instead of the right block
or we can put the vertex instead of the left block and
$\g_2$ in place of the right block.
The result is
\bea
F_{\g_3}(\eta)
&=&\int_0^\eta F_{\g_2}(x,\,x,\,x)F_{\g_0}dx
+\int_0^\eta F_{\g_0}F_{\g_2}(x,\,x,\,x)dx\\
&=&2\int_0^\eta \frac{1}{2}x^2dx
=\frac{1}{3}\eta^3.
\eea
This result should be compared with (\ref{ll3}).

\section{Two charge model}

In this section, we make an attempt to find an interpretation of
the new RG equation.
Let us first revisit the logic of derivation of leading RG
logarithms in QFT.
During the subtraction of divergences, one has to
introduce some scale necessary for the definition of the theory.
Physical observables should not depend on this scale.
This is achieved by the introduction of running couplings that
depend on the scale in such a way that the observables are
invariant.
The possibility to do so imposes some restrictions on the theory.
One of the consequences of these restrictions is that the leading
logarithms are defined by the one-loop beta function.
Thus the
renormalizability of the theory fixes the form of the
leading logarithms.

The question is whether the RG invariance of observables imposes
some conditions on individual Feynman graphs.
In a given process, only a sum of Feynman integrals is the
observable amplitude.
But in different processes, we may have different combinations of
the integrals,
and this may impose some constrains on the Feynman integrals.
For example, consider the two-charge model
\be
L(\vp,\chi)=\frac{1}{2}(\p\vp)^2+\frac{1}{2}(\p\chi)^2
+ \frac{g}{4!}\:\vp^4 + \frac{\td{g}}{4}\:\vp^2\chi^2.
\ee
We will study two processes:
$2\vp\ra 2\vp$ (two $\vp$ particles in two $\vp$ particles) and
$2\chi\ra 2\vp$ (two $\chi$ particles in two $\vp$ particles).
Assume that $\td{g}<<g$. Then, in the second process, we neglect
$\td{g}^2$.
Additionally, we let the external momenta be almost the same, i.e. we will
not study the non RG logarithms.
In two-loop approximation we have
\bea
F(2\vp\ra 2\vp)
&=&g-\frac{3}{2}g^2F(\g_1)+\frac{3}{4}g^3 F(\g_{11})
+\frac{6}{2}g^3 F(\g_{2});\\
F(2\chi\ra 2\vp)
&=&\td{g}-\frac{1}{2}\td{g}g F(\g_1)+\frac{1}{4}\td{g}g^2 F(\g_{11})
+\frac{1}{2}\td{g}g^2 F(\g_{2}).
\eea
In the second amplitude, we neglect the diagrams in $t$ and
$u$ channels since they are proportional to $\td{g}^2$.
The diagram $\g_2$ has only one orientation (out of 6 possible).
The RG equations for the couplings are
\beaa\nonumber
\frac{\p}{\p\log\Ld^2}g&=&\frac{3}{2}g^2;\\
\lb{beta}
\frac{\p}{\p\log\Ld^2}\td{g}&=&\frac{1}{2}g\td{g}.
\eeaa
In the leading logarithm approximation, we have
\beaa\nonumber
\frac{d}{d\log\Ld^2}F(2\vp\ra 2\vp)
&=&0;\\
\lb{sys}
\frac{d}{d\log\Ld^2}F(2\chi\ra 2\vp)
&=&0.
\eeaa
Taking into account equations (\ref{beta}),
we rewrite the system (\ref{sys}) as
\bea
\p_\tau F(\g_{11})+2\p_\tau F(\g_{2})
&=&4F(\g_1);\\
\p_\tau F(\g_{11})+4\p_\tau F(\g_{2})
&=&6F(\g_1),
\eea
where $\tau=\log\Ld^2$.
The solution of this system is
\bea
\p_\tau F(\g_{11})&=&2F(\g_1);\\
\p_\tau F(\g_{2})&=&F(\g_1).
\eea
We see that in the two charge model, the RG equations for the
two-loop Feynman integrals take the form of Connes and Kreimer RG
equations.
This example suggests the following interpretation of RG equations
(\ref{rgmain}).
These RG equations ensure that, in any theory with several couplings
and particles, the amplitudes will be RG invariant.
Also it is plausible that the inverse statement is true,
and there exists a theory with an infinite number of charges and
particles such that the RG invariance of all the amplitudes in
this theory is equivalent to RG equations
(\ref{rgmain}).

\bigskip

{\bf Conclusion.}
In the paper, we have studied the RG equations for individual Feynman
integrals.
We argue that the RG invariance of any renormalizable
theory should follow from
equations of this type or, inversely,
that the new RG equations follow from the RG invariance of all
admissible theories.
Using these equations, we calculate complete leading logarithms in
$\Phi^4$ theory and compare the result with parquet approximation.

\bigskip
\bigskip

The author is thankful to K.Selivanov and V.Belokurov for constant attention
to the work, and to A.M.Polyakov for discussions and useful
comments.
The work is supported by RFBR grant 03-02-17373
and INTAS grant 00-334.


\begin{thebibliography}{40}

 \bibitem{Bog}
N.N.Bogoliubov, D.V.Shirkov,
Introduction to the Theory of Quantized Fields,
3rd ed., Wiley-Interscience, 1980.


 \bibitem{K1}
D. Kreimer,
Adv.Theor.Math.Phys. 2 (1998) 303-334,
 q-alg/9707029

 \bibitem{CK1}
 A.Connes, D.Kreimer,
Commun.\ Math.\ Phys.\  {\bf 210}, 249 (2000),\\
hep-th/9912092.


 \bibitem{CK2}
 A.Connes, D.Kreimer,
Commun.\ Math.\ Phys.\  {\bf 216}, 215 (2001),\\
 hep-th/0003188.



\bibitem{Ionescu:2003nm}
L.~M.~Ionescu and M.~Marsalli,
hep-th/0307112.


\bibitem{Connes:1998qv}
A.~Connes and D.~Kreimer,
Commun.\ Math.\ Phys.\  {\bf 199}, 203 (1998),\\
hep-th/9808042.





\bibitem{Kreimer:1999wh}
D.~Kreimer and R.~Delbourgo,
Phys.\ Rev.\ D {\bf 60}, 105025 (1999),\\
hep-th/9903249.

\bibitem{Broadhurst:1999ys}
D.~J.~Broadhurst and D.~Kreimer,
Phys.\ Lett.\ B {\bf 475}, 63 (2000),\\
hep-th/9912093.



 \bibitem{GMS}
 A.Gerasimov, A.Morozov, K.Selivanov,
 Int. J. Mod. Phys. A16: 1531 (2001),
 hep-th/0005053



\bibitem{ll4}
D.~Malyshev,
Phys.\ Lett.\ B {\bf 578}, 231 (2004),
hep-th/0307301.



\bibitem{dyatlov}
I.~T.~Dyatlov, V.~V.~Sudakov, and K.~A.~Ter-Martirosyan,
Sov.\ Phys.\ JETP {\bf 5}, 631 (1957),
[Zh.\ Eksp.\ Teor.\ Fiz.\  {\bf 32}, 767 (1957)].


\bibitem{polyakov}
A.~M.~Polyakov,
Sov.\ Phys.\ JETP {\bf 30}, 151 (1970),
[Zh.\ Eksp.\ Teor.\ Fiz.\  {\bf 57}, 271 (1969)].


 \end{thebibliography}
\end{document}